\documentclass[letterpaper,10pt,conference]{ieeeconf}
\pdfoutput=1

\usepackage{color}
\usepackage[dvipsnames]{xcolor}
\usepackage{amsmath}
\usepackage{amssymb}
\usepackage{graphicx}
\usepackage{bm}
\usepackage{url}
\usepackage[font=footnotesize, skip=4pt]{caption}
\usepackage{subfig}
\IEEEoverridecommandlockouts
\newcommand{\Ncal}{{\mathcal{N}}}

\makeatletter
\newcommand{\revisionhistory}[1]{%
\@ifundefined{showrevisionhistory}{\relax}{%
{#1}%
}%
}
\makeatother

\begin{document}

\title{\LARGE \bf Residential Demand Response Targeting\\ Using Machine Learning with Observational Data}
\author{Datong Zhou, Maximilian Balandat, and Claire Tomlin\thanks{Datong Zhou is with the Department of Mechanical Engineering, University of California, Berkeley, USA. {\tt\footnotesize datong.zhou@berkeley.edu}}
\thanks{Maximilian Balandat and Claire Tomlin are with the Department of Electrical Engineering and Computer Sciences, University of California, Berkeley, USA.
{\tt\footnotesize [balandat,tomlin]@eecs.berkeley.edu}}%
\thanks{This work has been supported in part by the National Science Foundation under CPS:FORCES (CNS-1239166).}
\thanks{Datong Zhou is funded by the Berkeley Fellowship for Graduate Study.}}

\maketitle
\thispagestyle{empty}
\pagestyle{empty}

\begin{abstract}
The large-scale deployment of Advanced Metering Infrastructure among residential energy customers has served as a boon for energy systems research relying on granular consumption data. 
Residential Demand Response aims to utilize the flexibility of consumers to reduce their energy usage during times when the grid is strained. Suitable incentive mechanisms to encourage customers to deviate from their usual behavior have to be implemented to correctly control the bids into the wholesale electricity market as a Demand Response provider. In this paper, we present a framework for short-term load forecasting on an individual user level, and relate non-experimental estimates of Demand Response efficacy (the estimated reduction of consumption during Demand Response events) to the variability of a user's consumption. We apply our framework on a dataset from a residential Demand Response program in the Western United States. Our results suggest that users with more variable consumption patterns are more likely to reduce their consumption compared to users with a more regular consumption behavior.
\end{abstract}


%
%


\section{Introduction}
\label{sec:Introduction}

The widespread deployment of Advanced Metering Infrastructure (AMI) has made granular data on the electricity consumption of individual residential electricity customers available on a large scale. Smart meters report the electricity consumption of customers at a high temporal resolution, which enables novel data-centric services. One such service is residential Demand Response (DR), in which a DR provider serves as an interface between individual residential customers and the wholesale electricity market. The economic argument made for DR is that it is believed to improve economic efficiency by providing program participants with a proxy of a price signal~\cite{Commission:2010aa}.

Regulators and market operators in different jurisdictions have been moving towards allowing DR providers to offer capacity directly into wholesale electricity markets~\cite{PJM:2014aa:CapPerf,CPUC:2015:E4728}. The DR provider incentivizes users to temporarily reduce consumption at certain times, e.g. during periods of high Locational Marginal Prices (LMPs), bundles these reductions, and makes capacity bids into the market. If dispatched, the DR provider has to provide a reduction in energy consumption with respect to a certain baseline, and is rewarded by the LMP at the time of dispatch (in this paper we are not concerned with the capacity payments that DR providers receive for helping to fulfill Resource Adequacy requirements).  
In such auction-based market settings, it is crucial for DR providers to be able to make informed bids, as bidding too much capacity might result in a penalty due to failure to meet obligations, and bidding too little would result in a suboptimal revenue. The process of making bids is a complex problem - factors to take into account are, among others, the LMP, which determines the marginal price for DR reductions, the number of responsive DR participants under contract, and some knowledge about the behavior of these participants during DR event periods. The DR provider can improve its bidding strategy and efficiency by modeling the users' consumption behavior during DR and non-DR periods and by targeting households with a high potential reduction during DR hours. 

In this paper, we identify such users through a combination of established Machine Learning (ML) methods for short-term load forecasting (STLF), load shape clustering, and non-parametric statistics. STLF is employed to predict the consumption of individual residential customers during regular operation as well as during DR periods. This is used in conjunction with a non-parametric hypothesis test to determine whether, under our modeling assumptions, a reduction of consumption during DR periods can be detected~\cite{Balandat:2016aa}. These reductions serve as non-experimental estimates of participants' willingness to reduce energy consumption during DR periods. We then identify a ``dictionary" of consumption patterns by clustering load shapes in order to correlate the variability of an individual's consumption pattern to our non-experimental estimate of their consumption shift. Our results show a positive correlation between the degree of variability and our non-experimental estimates of the reductions. This finding may be used for adaptive targeting of users solely based on historical consumption data.

In the area of STLF, the two main categories of research are statistical time series modeling and techniques relying on predictor functions \cite{Munoz:2010aa}. The first category makes use of ARMA, ARIMA, and SARIMA models \cite{Taylor:2007aa, Arora:2014aa}, and the second uses classical regression techniques such as Least Squares, Lasso- and Ridge-Regression \cite{Fan:2012aa}, or a class of modern nonparametric methods in which Support Vector Regression \cite{Elattar:2010aa}, Nearest Neighbors Regression, Neural Networks \cite{Edwards:2012aa}, and fuzzy models \cite{Edwards:2012aa} have been most extensively studied. Other approaches are based on Principal Component Analysis \cite{Taylor:2007aa}, state-space models such as Kalman-Filtering \cite{Ghofrani:2011aa} or exponential smoothing methods (Holt-Winters Method) \cite{Mirowski:2014aa}. The covariates that are most often used for forecasting are previous observations of consumption, temperatures, and calendar variables such as hour of day and day of week.

Clustering algorithms have been investigated in \cite{Kwac:2014aa} and \cite{Smith:2012aa}, who conduct a segmentation of residential load shapes of consumption data in California. Similarly, \cite{Rhodes:2014aa} identifies typical load shapes of residential customers in Austin, Texas, and relates those to socioeconomic data. A comparison between common clustering algorithms - Hierarchical clustering, $k$-means clustering, and fuzzy $C$-means clustering - is performed in \cite{Kim:2011aa}. Other methods, including Self-Organizing Maps, are explored in \cite{Chicco:2012aa}.

The contribution of this paper lies in combining methods from STLF, load shape analysis, and non-parametric statistics to identify more responsive users for DR programs. The remainder of this paper is organized as follows: In Section \ref{sec:Preliminaries}, we introduce the data and outline preliminary steps. We describe ML algorithms used for STLF in Section \ref{sec:Forecasting_Techniques} and detail the estimation of energy reduction during DR hours in Section \ref{sec:Estimated_Treatment_Effects_DR}. In Section \ref{sec:Segmentation}, we present the methodology used for load shape analysis, and then apply our methods on both synthetic data (Section \ref{sec:Synthetic_Data}) and real consumption data (Section \ref{sec:Experiments_Data}). We conclude in Section \ref{sec:Conclusion}. 


%
%


\section{Preliminaries}
\label{sec:Preliminaries}

\subsection{Data Characteristics}
\label{sec:Data_Characteristics}
Our analyses are based on hourly smart meter readings of 500 residential electricity customers in the Western United States, collected between 2012 and 2014. Aligned with those readings are timestamps of notifications sent by the DR provider to the users that prompt them to reduce their consumption for a short period. We also use ambient air temperature measurements from public data sources to capture the correlation between temperature and electricity consumption.

\subsection{Data Preprocessing}
\label{sec:Data_Preprocessing}
Before any analysis is carried out, we pre-process the available data to provide a coherent basis for a comparison of different forecasting techniques.

First, we exclude users with residential solar photovoltaics to remove effects due to correlation in power generation and DR events. We also exclude users with corrupt meter readings (such as excessive or negative consumption). 

Second, the time series for consumption and temperature are matched by only taking data into account that includes both temperature and consumption readings. Temperature observations are resampled to hourly data by taking a weighted mean between non-evenly spaced measurements.

Third, consumption and temperature are standardized to zero mean and unit variance to allow future comparisons of prediction methods that are not necessarily scale-invariant.
 
Fourth, the consumption series are analyzed for stationarity with the augmented Dickey-Fuller test \cite{Fuller:1995aa}. In particular, it has to be asserted that DR events, interpreted as exogenous ``shocks", only have transitory effects and thus do not permanently impact the non-DR consumption. After differencing the consumption series in order to free it from seasonality, all the consumption time series are found to be stationary with a significance level of more then 99\%. This is in accordance with \cite{Mirowski:2014aa}, where the authors used the Kwiatkowiski, Phillips, Schmidt, and Shin Test (KPSS) to assert stationarity.

Fifth, and most importantly, hours of meter readings ``shortly" after DR hours are removed from the training dataset. For every DR message sent to a user, we remove 8 hours of subsequent metering recordings to prevent forecasting algorithms to use data in hours that are likely to have been influenced by users deviating from their usual consumption behavior. We therefore assume that the user reverts to the usual behavior at most 8 hours after receiving their last DR message. Most existing literature on the ``rebound effect",
which describes the increase of electricity consumption after the end of DR periods, is concerned with the consumption in a single hour after the DR event \cite{Mathieu:2011aa, Li:2012aa}. Thus, removing 8 hours is a conservative estimate to remove spillovers of consumption anomalies into the training data.

\subsection{Covariates}
\label{sec:Covariates}
Our models for participants' consumption are fitted by using the following predictors:
\begin{itemize}
\item Previous hourly consumptions,
\item Previous hourly ambient temperatures,
\item A categorical variable combining the hour of day with a boolean weekend indicator variable.
\end{itemize}
That is, to predict the consumption at a given time, we incorporate an autoregressive term taking into account five previous meter readings, five past ambient temperatures as well as a categorical variable of length 48 that differentiates between a weekday/weekend day and the hour of day.

\subsection{Data Splitting}
\label{sec:Data_Splitting}
The pre-processed data is split into a training set that represents the ``usual" behavior of users during non-DR hours, and a DR set that describes the consumption of users during DR events. The outcome/covariate pairs for a given user $i$ are denoted as $\left( Y_i^0, X_i^0 \right)$ and $\left( Y_i^1, X_i^1 \right)$ for the training set and the DR set, respectively.

%
%


\section{Forecasting Techniques}
\label{sec:Forecasting_Techniques}
We apply the following forecasting methods:
\begin{itemize}
\item \label{itm:LS} Ordinary Least Squares Regression (OLS)
\item \label{itm:Lasso} Lasso Regression (L1)
\item \label{itm:Ridge} Ridge Regression (L2)
\item \label{itm:KNN} $k$ Nearest Neighbors Regression (KNN)
\item \label{itm:SVR} Support Vector Regression (SVR)
\item \label{itm:DT} Decision Tree Regression (DT)
\item ISO Baseline Prediction
\end{itemize}
Each forecasting model, trained on $\left( Y_i^0, X_i^0 \right)$, is then applied to the covariates of the DR data $X_i^1$ to obtain the estimated consumption $\hat{Y}_i^c$. 
This prediction is subsequently compared to the observed consumption $Y_i^1$ during DR events. The differences
\begin{equation}
Y_i^{\Delta} = Y_i^1 - \hat{Y}_i^c
\end{equation}
will be used to compare the statistical differences between consumption predictions outside and during DR periods.

\subsection{Ordinary Least Squares Regression}
\label{sec:LS_Regression}
Assuming a linear relationship between covariate-outcome pairs $(X_i^0(j), Y_i^0(j)),~j = 1, \ldots, N$,
\begin{equation}\label{eq:X_y_linear}
Y_i^0 = X_i^0 \beta,
\end{equation}
the parameter $\beta$ is estimated using OLS.

\subsection{Ridge- and Lasso-Regression}
\label{sec:L2_L1}
The same linear relationship \eqref{eq:X_y_linear} with a regularization term on the parameter $\beta$ results in 
LASSO-Regression 
\begin{equation}\label{eq:Lasso}
\hat{\beta} = \arg\min_{\beta} \| Y_i^0 - X_i^0 \beta \|_2^2 + \lambda \| \beta \|_1
\end{equation}
or Ridge-Regression
\begin{equation}\label{eq:Ridge}
\hat{\beta} = \arg\min_{\beta} \| Y_i^0 - X_i^0 \beta \|_2^2 + \lambda \| \beta \|_2.
\end{equation}
The penalty $\lambda$ is found with standard cross-validation techniques (e.g. we used $k$-fold cross-validation) \cite{Hastie:2013aa}.

\subsection{KNN-Regression}
\label{sec:KNNR}
Given a point in feature space $X_i^0(z)$, the goal is to find the $k$ training points $X_i^0(1), \ldots, X_i^0(k)$ that are closest in distance to $X_i^0(z)$ \cite{Hastie:2013aa}. We chose the commonly used Euclidian norm (though other choices can be justified) as a measure for distance in feature space.
The prediction of the outcome variable $\hat{Y}_i^0(z)$ is the average of the outcomes of the $k$ nearest neighbors
\begin{equation}
\hat{Y}_i^0(z) = \frac{1}{k}\left[ Y_i^0(1) + \ldots + Y_i^0(k) \right].
\end{equation}
The number of neighbors $k$ is found using cross-validation to avoid overfitting ($k$ too small) and underfitting ($k$ too large).

\subsection{Support Vector Regression}
\label{sec:SVR}
Support Vector Regression solves the following optimization problem:
\begin{equation}\label{eq:SVR_opt}
\begin{aligned}
&\min_{w, b, \xi, \xi^{\ast}} \frac{1}{2} \|w\|^2 + C\sum_{j=1}^{N}(\xi_j + \xi_j^{\ast})\\
\text{s.t.}~ &Y_i^0(j) - w^{\top} \phi(X_i^0(j)) - b \leq \epsilon +\xi_j,\\
&w^{\top} \phi(X_i^0(j)) + b - Y_i^0(j) \leq \epsilon +\xi_j^{\ast},\\
&\xi_j, \xi_j^\ast \geq 0 \quad \forall j \in [1, \ldots, N].
\end{aligned}
\end{equation}
In \eqref{eq:SVR_opt}, $\epsilon$ defines an error tube within which no penalty is associated, $\xi$ and $\xi^{\ast}$ denote slack variables that guarantee the existence of a solution for all $\epsilon$, $b$ is a real constant, $C$ is the regularization constant, $w$ are the regression coefficients to be estimated, and $\phi(\cdot)$ a map between the input space and a higher dimensional feature space. \eqref{eq:SVR_opt} is typically solved by transforming it into dual form, thereby avoiding the explicit calculation of $\phi(\cdot)$ with the so-called Kernel trick. We chose the commonly used Gaussian Kernel function. The resulting optimization problem can be readily solved using Quadratic Programming \cite{Elattar:2010aa}.

\subsection{Decision Tree Regression}
\label{sec:DTR}
This non-parametric learning method finds decision rules that partition the feature space into up to $2^n$ pieces, where $n$ denotes the maximal depth of the tree. For a given iteration step, enumeration of all nodes and possible splitting scenarios (exhaustive search) yields a tuple $\theta^{\ast} = (j, t_m)$ that minimizes the sum of the ensuing child node impurities $G(\theta^{\ast}, m)$, where $j$ denotes the $j$-th feature and $m$ the $m$-th node of the tree. This is formally written as
\begin{align}\label{eq:DT_optimization}
\theta^{\ast} &= \arg\min_{\theta} G(\theta, m), \\
\label{eq:DT_optimization_2} G(\theta, m) &= \frac{n_{\text{left}}^m}{N_m}H(Q_{\text{left}}(\theta)) + \frac{n_{\text{right}}^m}{N_m}H(Q_{\text{right}}(\theta)).
\end{align}
where $Q_{\text{left}}$ and $Q_{\text{right}}$ denote the set of $\left( X_i^0, Y_i^0 \right)$ covariate-outcome pairs belonging to the left and right child node of mother node $m$, respectively; and $n_{\text{left}}^m$ and $n_{\text{right}}^m$ denote their respective count. The impurity measure $H(\cdot)$ at a node minimizes the mean squared error
\begin{align}\label{eq:impurity_MSE}
c(\cdot) &= \frac{1}{N(\cdot)} \sum_{j \in N(\cdot)} Y_i^0(j), \\
\label{eq:impurity_MSE_2} H(\cdot) &= \frac{1}{N(\cdot)} \sum_{j \in N(\cdot)} \left[Y_i^0(j) - c(\cdot)\right]^2,
\end{align}
with $N(\cdot)$ representing the number of observations at the node of interest.

One popular algorithm is exhaustive search, e.g. \textit{CART}, which iteratively performs \eqref{eq:DT_optimization}, \eqref{eq:DT_optimization_2}, \eqref{eq:impurity_MSE}, and \eqref{eq:impurity_MSE_2} to detect the optimal tuple $\theta^{\ast} = (j, t_m)$ to update the tree until some convergence criterion, e.g. on the maximal depth, is reached~\cite{Breiman:1984aa}.
%
Cross-validation, usually on the number of the maximal depth of the tree (which we used) or the minimal number of samples per node, avoids overfitting of the tree. Novel cross-validation schemes on Decision Trees specifically tailored to estimate the outcome of treatment effects have been proposed by \cite{Athey:aa}.
The optimized tree is then used for forecasting the outcome by taking the average of all outcomes belonging to a given node $m$. This yields a decision tree with piecewise constant predictions.

\subsection{ISO Baseline Prediction}
\label{sec:CAISO_Baseline}
For benchmarking purposes, we use a baselining procedure as commonly employed by many Independent System Operators (ISOs). Techniques vary between ISOs, but yield similar results~\cite{KEMA:2011aa:Baselines}. We chose the so-called ``10 in 10'' methodology as defined by the California Independent System Operator (CAISO)~\cite{CAISO:2014aa:Tariff}. 
The baseline for a given hour on a weekday is obtained by averaging the hourly consumption during the given hour of the past 10 weekday consumptions on days without an event. Similarly, the baseline for a given hour on a weekend day or holiday is the average consumption during the given hour observed on 4 past weekend days or holidays~\cite{CAISO:2014aa:Tariff}. Further, the baseline on a day of a DR event is modified with a so-called \textit{Load Point Adjustment}
by multiplying the hourly baseline values with a ratio, which is calculated as the mean consumption of the three hours preceding the hour before the DR event compared to the average baseline for the same hours.

%
%


\section{Non-Experimental Estimates of DR Treatment Effects}
\label{sec:Estimated_Treatment_Effects_DR}

\subsection{Counterfactual DR Consumption}
\label{sec:Counterfactual_Consumption}
Following the general idea of~\cite{Balandat:2016aa}, we use the different models fitted on the training data to obtain a non-experimental estimate of the \textit{counterfactual} consumption $\hat{Y}_i^c$, which can be described as the \textit{consumption during DR times in the hypothetical absence of a DR event}. This consumption of course cannot be observed on the level of an individual, since at all DR times, the DR event has affected the consumption of a given user. This general problem has been referred to as the fundamental problem of causal inference~\cite{Holland:1986aa}. Since model misspecification cannot be ruled out, any true causal estimate of treatment effects will require the comparison of different groups in a randomized controlled experiment. However, conducting such an experiment typically involves significant preparation time and cost. The contribution of our approach is that it allows to generate meaningful non-experimental estimates in a much broader range of settings. 

As a proxy for the unobservable counterfactual consumption in the absence of a DR event, we use the prediction $\hat{Y}_i^c$ obtained by the cross-validated forecasting techniques presented earlier. 
We define the average empirical reduction $\hat{\Delta}_i$ for user $i$ during DR hours as 
\begin{equation}\label{eq:est_treatment_eff}
\hat{\Delta}_i = \frac{1}{N} \sum_{j=1}^N \bigl( \hat{Y}_i^c(j) - Y_i^1(j) \bigr),
\end{equation}
which is simply the sample mean of the component-wise difference between the estimated counterfactual and the actual, observed DR consumption. $N$ represents the number of DR events. The intuition is that the forecasting models have been trained on non-DR data $\left(X_i^0, Y_i^0\right)$, and predictions for DR consumptions $\hat{Y}_i^c$ assume the absence of DR events. Therefore, if the mean of the estimated counterfactual consumption exceeds the mean of the actual DR consumption $Y_i^1$, then, assuming the absence of model mismatch, the difference in means can be interpreted as the mean reduction during DR events. Note that $\hat{\Delta}_i$ is not restricted to positive values - according to \eqref{eq:est_treatment_eff}, a negative $\hat{\Delta}_i$ would imply an increased DR consumption by a mean value of $|\hat{\Delta}_i|$.

Equation \eqref{eq:est_treatment_eff} is an absolute measure that ignores the respective overall consumption level. For a potentially more meaningful, relative measure, we define the \textit{weighted mean percentage reduction} (MPR)
\begin{equation}\label{eq:MPR}
\text{MPR} = \frac{1}{N}\sum_{j=1}^N \frac{Y_i^1(j) - \hat{Y}_i^c(j)}{|\hat{Y}_i^c(j)|}\cdot 100\%,
\end{equation}
which normalizes the componentwise deviations by the estimated counterfactual consumption. $\text{MPR} < 0 $ corresponds to an estimated average DR reduction of $|\text{MPR}| \%$. Note that a disadvantage of MPR lies in the normalization of the componentwise deviations by $|\hat{Y}_i^c(j)|$, which gives disproportionate errors for small $|\hat{Y}_i^c(j)|$.

\subsection{Nonparametric Hypothesis Test}
\label{sec:Nonparametric_Hypothesis_Test}
$\hat{\Delta}_i$ and $\text{MPR}$ can be evaluated on a set of DR events belonging to an individual user to estimate individual treatment effects, or an aggregation of users to estimate average treatment effects. Clearly, the accuracy of the estimated average treatment effects scales with the size of the user base (modulo potentially unmodeled effects).

However, $\hat{\Delta}_i$ and $\text{MPR}$ on an individual user level will typically be very noisy due to the volatility of the consumption behavior of a single user. Therefore, we make use of a nonparametric hypothesis test to compare our estimates on individual users, following the approach presented in \cite{Balandat:2016aa}. This is carried out by comparing the samples $Y_i^1$ and $\hat{Y}_i^c$ with the Wilcoxon Signed Rank Test, whose goal is to determine that these samples stem from different distributions. Our null hypothesis is that both samples are generated by the same (unknown) distribution $F(u)$:
\begin{equation}\label{eq:null_hypothesis_samples}
H_0: Y_i^1, \hat{Y}_i^c \sim F(u) \Rightarrow \mathbb{E}\bigl[Y_i^1 - \hat{Y}_i^c \bigr] = 0.
\end{equation}
The null hypothesis \eqref{eq:null_hypothesis_samples} is juxtaposed with the (one-sided) alternative hypothesis $H_1$, stating the existence of a difference in the distribution between $Y_i^1$ and $\hat{Y}_i^c$:
\begin{equation}\label{eq:alternative_hypothesis}
H_1: Y_i^1 \sim F(u),~ \hat{Y}_i^c \sim F(u) + \Delta \Rightarrow \mathbb{E}\bigr[\hat{Y}_i^c - Y_i^1\bigr] = \Delta.
\end{equation}
In \eqref{eq:alternative_hypothesis}, it is assumed that the predicted outcomes $\hat{Y}_i^1$ are generated by a distribution of the same shape, but shifted by the parameter $\Delta$:
\begin{equation}
\mathbb{E}\bigl[\hat{Y}_i^c - Y_i^1\bigr] = \Delta.
\end{equation}
If the alternative hypothesis is accepted, this suggests that, within the constraints of our model, the predicted counterfactual consumptions $\hat{Y}_i^c$ are on average greater than the observed DR consumptions $Y_i^1$, which can be interpreted as a mean reduction of consumption by $\Delta$ during DR hours. Further, the $p-$value of the hypothesis test is the probability of making the observations under the null hypothesis.


\subsection{Wilcoxon Signed Rank Test}
\label{sec:Wilcoxon}
The Wilcoxon Signed Rank Test with paired samples $(\hat{Y}_i^c, Y_i^1)$ is a non-parametric hypothesis test that follows the intuition of the paired Student's $t$-test, but does not assume that the samples are drawn from a normal distribution. The test places the pairwise sample differences, excluding zero differences, into a single list which is then sorted in ascending order. Next, the rank of a data point is defined as its ordinal position in this sorted list. The test statistic $U$ involves the sum of the signed ranks, which for large sample sizes can be approximated by a normal distribution:
\begin{align}
U\sim\Ncal\left(\mu,\sigma^2\right),~ \mu = 0,~\sigma^2 = \frac{N(N+1)(2N+1)}{6},
\end{align}
where $N$ denotes the number of non-identical paired samples in $(\hat{Y}_i^c, Y_i^1)$.
In addition to the p-value, an estimate $\hat{\Delta}$ of the location parameter shift can be obtained via the Wilcoxon Signed Rank Test (the so-called Hodges-Lehmann estimator). This estimate corresponds to the mean empirical reduction of consumption during DR-events of user $i$ based on the samples $(\hat{Y}_i^c, Y_i^1)$ \cite{Hettmansperger:2011aa}. 

%
%


\section{Segmentation of Users}
\label{sec:Segmentation}
The consumption behavior of residential electricity customers is highly variable across the population, and many analyses have been performed on the relationship between socioeconomic factors and household energy consumption (see e.g. \cite{Yun:2011aa, Bhattacharjee:2011aa}). 
Inspired by these approaches, we explore the existence of a relationship between the \textit{variability} of user consumption and our non-experimental estimates of the change in consumption during DR periods. Any conclusion drawn from this analysis would be useful for the purpose of targeting particular consumers and allow for a more efficient identification and recruiting of users for DR programs.

\subsection{Load Shape Analysis}
\label{sec:Load_Shape_Analysis}
The intuition is to find a set of representative ``signature" load shapes that describe the user behavior. In other words, among the set of all observed load shapes, it is desirable to find a reduced set of load shapes that best describes consumption patterns. For this purpose, following \cite{Kwac:2014aa}, we define a load shape $s(t)$ consisting of 24 hourly values as 
\begin{equation}\label{load_shape}
a = \sum_{t=1}^{24}l(t) ~\text{and}~ s(t) = \frac{l(t)}{a},
\end{equation}
where $l(t)$ is a daily consumption profile $\in \mathbb{R}^{24}$ from midnight to midnight. We only collect weekday consumption patterns, as there is an increased variability of energy consumption during weekends \cite{Kwac:2014aa}. 
Next, in order to reduce the noise stemming from individual daily load shapes, for each user, 5 consecutive weekday load shapes are averaged \cite{Smith:2012aa} and treated as a single one. Denote the collection of all 5-day average loads as $\mathcal{S}$. 
Finding representative shapes $C_1, \ldots, C_k$ that minimize the squared error
\begin{equation}\label{eq:MSE_Clusters}
\text{SE}~ = \sum_{s_i \in \mathcal{S}} \left( C_i - s_i \right)^2,
\end{equation}
where $C_i$ denotes the cluster center closest to a given load shape $s_i$, is a clustering algorithm with $k$ clusters to be set.\\
Unlike \cite{Kwac:2014aa}, where the authors make use of a two-step $k$-means algorithm (first find the appropriate number of $k$ to maintain a maximal distance between loads and centers, then merge clusters together), we choose the standard $k$-means algorithm with different values of the number of a-priori defined cluster centers $k$.

\subsection{Variability of User Consumption}
\label{sec:Entropy}
After the $k$ cluster centers have been found, we characterize the \textit{variability} of a given user using the following metrics:
\subsubsection{Entropy}
Each daily load shape of user $j$ is matched to its closest cluster center. Define $p_j(C_i)$ as the frequency count of the event that a daily load shape is matched to centroid $i$ divided by the total number of load shapes. Then the entropy $H_j$ of user~$j$ is
\begin{equation}\label{eq:classical_entropy_def}
H_j = -\sum_{i=1}^k p_j(C_i)\log(p_j(C_i)).
\end{equation}
The entropy is minimal $(=0)$ if the user follows a single centroid, and maximal $(=\log(k))$ if all cluster centers are of equal occurrence \cite{Kwac:2014aa}.

\subsubsection{Hourly Standard Deviations}
We suggest the metric
\begin{equation}\label{eq:hourly_standard_dev}
\tilde{H}_j = \sum_{i=1}^{24} \text{std}\left[s_j(i)\right],
\end{equation}
i.e. the sum of the standard deviations of the observed hourly consumptions over all hours for a given user $j$. This method has the advantage that it does not rely on a $k$-means clustering algorithm, and thus avoids an a-priori choice of the number of clusters $k$.

%
%


\section{Validation on Synthetic Data}
\label{sec:Synthetic_Data}
We now construct synthetic data to verify the functionality of our forecasting algorithms and predicted counterfactual consumption to estimate the DR reduction. Our motivation is that we can benchmark our models on the a-priori known ground truth of the synthetic data. The goal is to show that, within the limitations of our model, our learning algorithms are capable of predicting the average empirical reduction \eqref{eq:est_treatment_eff} and the MPR \eqref{eq:MPR} with acceptable accuracy.

To generate an artificial time series $\bar{l}(t)$, a base consumption consisting of the daily characteristic load shapes shown in Figure \ref{fig:twelve_load_shapes} is constructed. The relative occurrence of the 12 dictionary load shapes in the base consumption is varied so as to generate timeseries with different entropies \eqref{eq:classical_entropy_def}. Then, a linear temperature contribution is added as well as Gaussian Noise $\epsilon \sim \mathcal{N}\left(0, \sigma^2\right)$. Further, a random subset of the time indices are defined as DR hours, for which the respective consumption is decreased by a constant $c_{\text{DR}} > 0$. The resulting artifical load shape $\bar{l}(t)$ therefore includes (known) components of the daily characteristic load shapes, the ambient temperature, and DR reductions:
\begin{equation}\label{eq:Synthetic_Consumption}
\bar{l}(t) = C_i(t) + c_t\cdot T(t) - \mathbb{I}(t\in \mathcal{D})\cdot c_{\text{DR}} + \epsilon(t),
\end{equation}
where $\mathcal{D}$ denotes the set of all DR times, $C_i(t)$ the cluster center in the base consumption at time $t$, and $c_t$ the proportionality constant for the ambient temperature at time $t$. After $\bar{l}(t)$ is standardized, we can run our forecasting techniques on this artificial load shape with the same features used in Section \ref{sec:Covariates}, and investigate the prediction accuracy as well as the estimates of the DR reductions as a function of the entropy and magnitude of noise.

\begin{figure}[hbtp]
\centering
\vspace*{-0.95cm}
\hspace*{-0.72cm}
\includegraphics[scale=0.33]{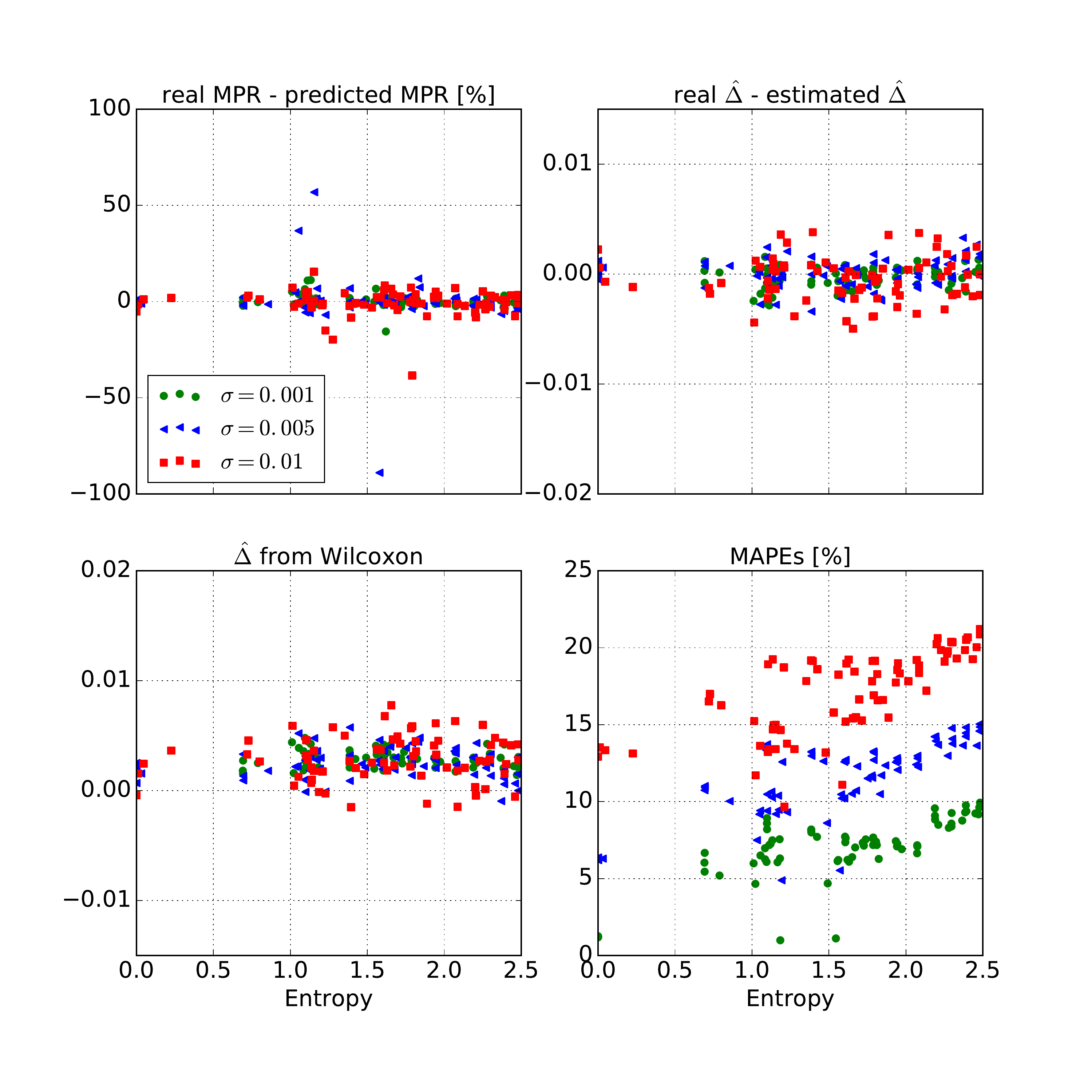}
\vspace*{-1.0cm}
\caption{Synthetic Data Characteristics with Different Noise Levels. Top Left: Actual MPR $-$ Predicted MPR, Top Right: Actual $\hat{\Delta}-$ Predicted $\hat{\Delta}$, Bottom Left: Wilcoxon-$\hat{\Delta}$, Bottom Right: MAPEs}
\label{fig:Synthetic_Data}
\vspace*{-0.0cm}
\end{figure}

Figure \ref{fig:Synthetic_Data} shows scatter plots for three different noise levels $\sigma$ estimated with Ridge-Regression. The plot shows the differences between actual and predicted MPR \eqref{eq:MPR}, the differences between the known mean reduction and the estimated mean reduction \eqref{eq:est_treatment_eff}, the estimated location parameter shift $\hat{\Delta}$ from the Wilcoxon Signed Rank Test, and the mean absolute percentage error of the consumption predictions (MAPE, \eqref{eq:MAPE}). Subplots 1-2 indicate that higher noise levels do not qualitatively impact the accuracy of prediction for MPR and the empirical reduction, even though the range of errors increases as $\sigma$ increases. Similarly, the estimated location parameter shift $\hat{\Delta}$ from the Wilcoxon Signed Rank Test varies around a constant, which, from further analyses, is found to be $c_{\text{DR}}$. As expected, higher noise levels increase the MAPE of the predictions. The observations imply that, under the correct model specification and in the absence of confounding variables, Ridge-Regression on average is able to correctly estimate the MPR, the empirical reduction, and the subsequent location parameter shift given by the Wilcoxon Signed Rank test, even in the presence of varying noise levels. It is also important to note that the findings of Subplots 1-3 are independent of entropy. Only subplot 4 shows an increase of MAPE as entropy increases, which intuitively makes sense because more variable consumption is inherently harder to predict. Lastly, further analyses show that the qualitative nature of Figure \ref{fig:Synthetic_Data} varies with the bias of the estimator, in the sense that upward biased estimates yield a higher $\hat{\Delta}$. A more precise analysis of the bias-variance tradeoff that influences the non-experimental estimates of $\hat{\Delta}$ is performed in \cite{Balandat:2016aa}.

We can think of real load shapes as some mixture of base load shapes, which describe the daily behavior of users. These base loads are then perturbed with temperature influences (e.g. increased AC consumption during high temperature days) and noise (e.g. user vagaries). It can be imagined that different users possess different archetypes of energy consumption behavior (e.g. a single person household might have a more regular consumption pattern than a family), and thus different entropies. Since our analysis on the synthetic data shows that the mean predicted DR reductions are independent of entropy, we conclude that our prediction algorithms are applicable to participants with different levels of consumption variability.

%
%


\section{Experiments on Data}
\label{sec:Experiments_Data}

\subsection{Prediction Accuracy}
\label{sec:Prediction_Accuracy}
\begin{figure}[hbtp]
\vspace*{-0.45cm}
\centering
\includegraphics[scale=0.36]{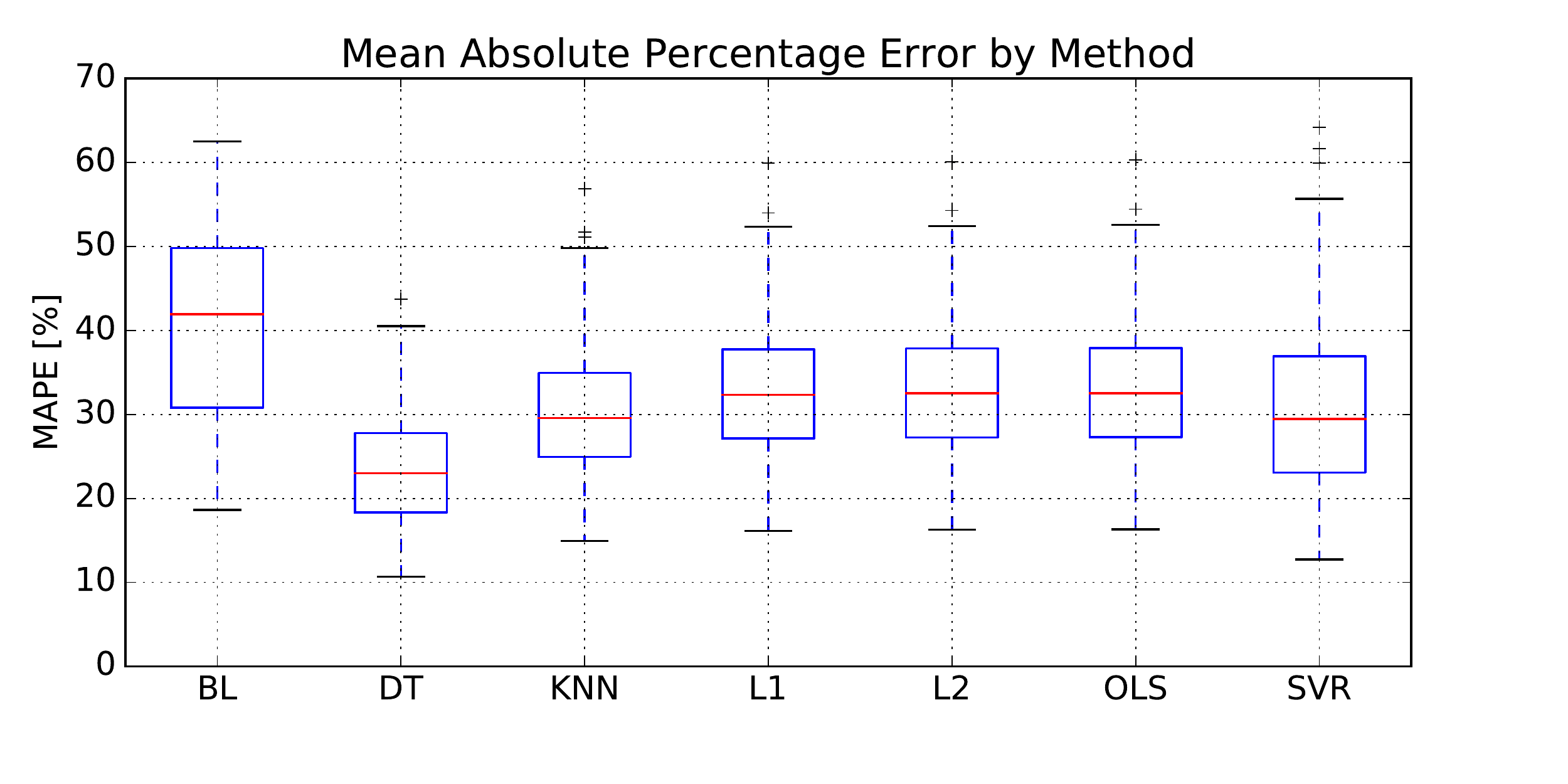}
\vspace*{-0.75cm}
\caption{MAPEs for Different Forecasting Techniques and CAISO Baseline}
\label{fig:MAPEs}
\end{figure}

We chose the mean absolute percentage error (MAPE) as a measure for the prediction accuracy:
\begin{equation}\label{eq:MAPE}
\text{MAPE}~=\frac{1}{N}\sum_{j=1}^N \frac{\vert Y_i^0(j) - \hat{Y}_i^0(j)\vert}{Y_i^0(j)}\cdot 100 \%.
\end{equation}
Figure \ref{fig:MAPEs} shows box plots for the MAPE of different prediction methods and the CAISO baseline across the user population. L1, L2 and LS have similar MAPEs, indicating that LASSO- and Ridge-Regression do not improve the prediction accuracy by adding bias and reducing variance. It is apparent that the L1-regularization term introduced in \eqref{eq:Lasso} does not have a strong effect, which indicates that because of the large data set available overfitting is not an issue. 
%
Similarly, the lack of MAPE reduction for Ridge-Regression indicates a lack of multicollinearity in the regressors \cite{Hettmansperger:2011aa}. As expected, the ISO baseline prediction performs worst since it averages hourly consumption readings far back in the past (up to 10 weekdays before a prediction), which are unlikely to predict the consumption accurately. Decision Trees and Support Vector Regression with median MAPEs of $\sim$ 23 and 29\%, respectively, outperform $k-$nearest neighbors and the linear regression methods whose median MAPE across users is $\sim 30-35 \%$. However, the better prediction quality of Support Vector Regression comes at a higher cost: Computation times of up to 45 minutes to fit an SVR model on a time series of length 40,000 were observed, whereas the training step for the linear regression models took less than 5 seconds per user on a six-core CPU. Prediction times for all methods, however, were negligible.

We acknowledge that more accurate predictions can likely be obtained by taking into account more covariates, e.g. a greater number of autoregressive consumption terms, more temperature data, and more sophisticated ML algorithms such as neural networks. This, however, is not the focus of this paper, and the reader is referred to \cite{Edwards:2012aa} for a discussion on the performance of forecasting algorithms.
\subsection{Estimation of Reduction of DR Consumption}
\label{sec:Reduction_DR_Consumption}
\begin{figure}[hbtp]
\vspace*{-0.35cm}
\centering
\includegraphics[scale=0.36]{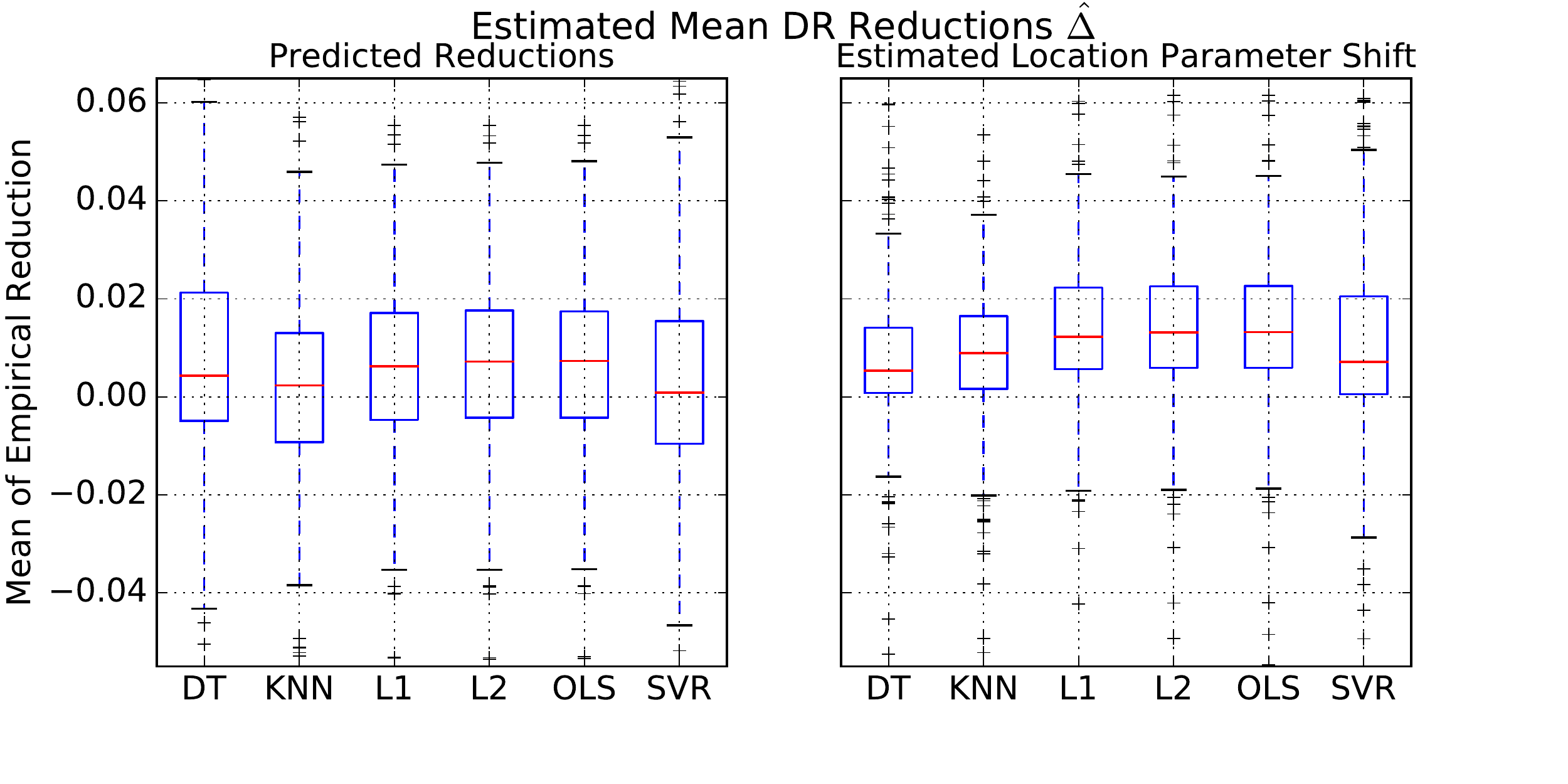}
\vspace*{-0.45cm}
\caption{Estimated DR Reductions. Left: Computed with \eqref{eq:est_treatment_eff}; Right: Wilcoxon-$\hat{\Delta}$}
\label{fig:Loc_Param_Shift}
\end{figure}

\begin{figure}[hbtp]
\vspace*{-0.35cm}
\centering
\includegraphics[scale=0.36]{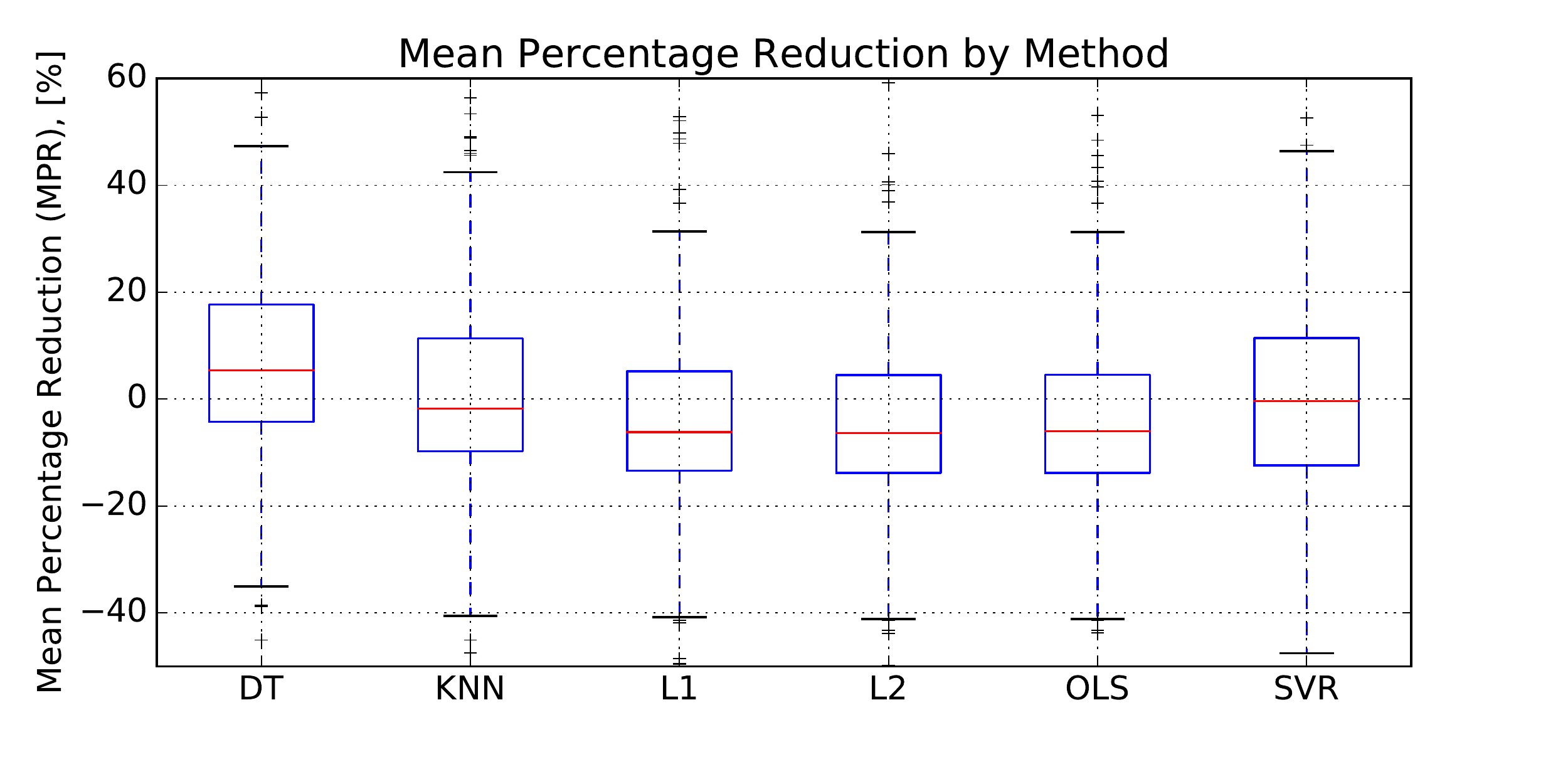}
\vspace*{-0.65cm}
\caption{Predicted Mean Percentage Reductions (MPRs)}
\label{fig:MPRs}
\end{figure}

Figure \ref{fig:Loc_Param_Shift} shows box plots of the estimated treatment effects determined by \eqref{eq:est_treatment_eff} and the location parameter shift estimates $\hat{\Delta}$ provided by the Wilcoxon Signed Rank Test, and Figure \ref{fig:MPRs} gives box plots of the range of estimated MPRs across all users by method, computed with \eqref{eq:MPR}. 

In Figure \ref{fig:Loc_Param_Shift} it can be seen that the median of the mean empirical reductions, computed with both the Wilcoxon Signed Rank Test and \eqref{eq:est_treatment_eff}, are greater than zero throughout. As already mentioned, the different levels of $\hat{\Delta}$ can be explained by potentially biased estimators, e.g. downward biased estimates of $\hat{Y}_i^c$, on average, yield a smaller $\hat{\Delta}$~\cite{Balandat:2016aa}. Indeed, our findings reveal that both KNN and SVR yield downward biased estimates across all users with a median value of 0.0025 and 0.0034, respectively. The bias for L1, L2 and DT was found to be less than $1e-09$ for all users. This explains the smaller median $\hat{\Delta}$ for KNN and SVR.

According to Figure \ref{fig:MPRs}, the median MPRs are between $\approx -0.2\%$ and $-7\%$ for all methods excluding DT, which is synonymous with a DR reduction in all cases but DT. It can be seen that the downward biased methods SVR and KNN result in a smaller median reduction $|\text{MPR}|$.
For DT, the counterintuitive result of an increased DR consumption (MPR $>0$) despite a near zero bias could possibly be explained with the normalization of some $\bigl(Y_i^1(j)-\hat{Y}_i^1(j) \bigr)$ by outliers in $|\hat{Y}_i^1(j)|$ that are close to zero because of misclassifications in the training step. 


\subsection{K-Means Clustering Results}
\label{sec:kmeans_clustering_results}
\begin{figure}[hbtp]
\centering
\vspace*{-0.35cm}
\includegraphics[scale=0.23]{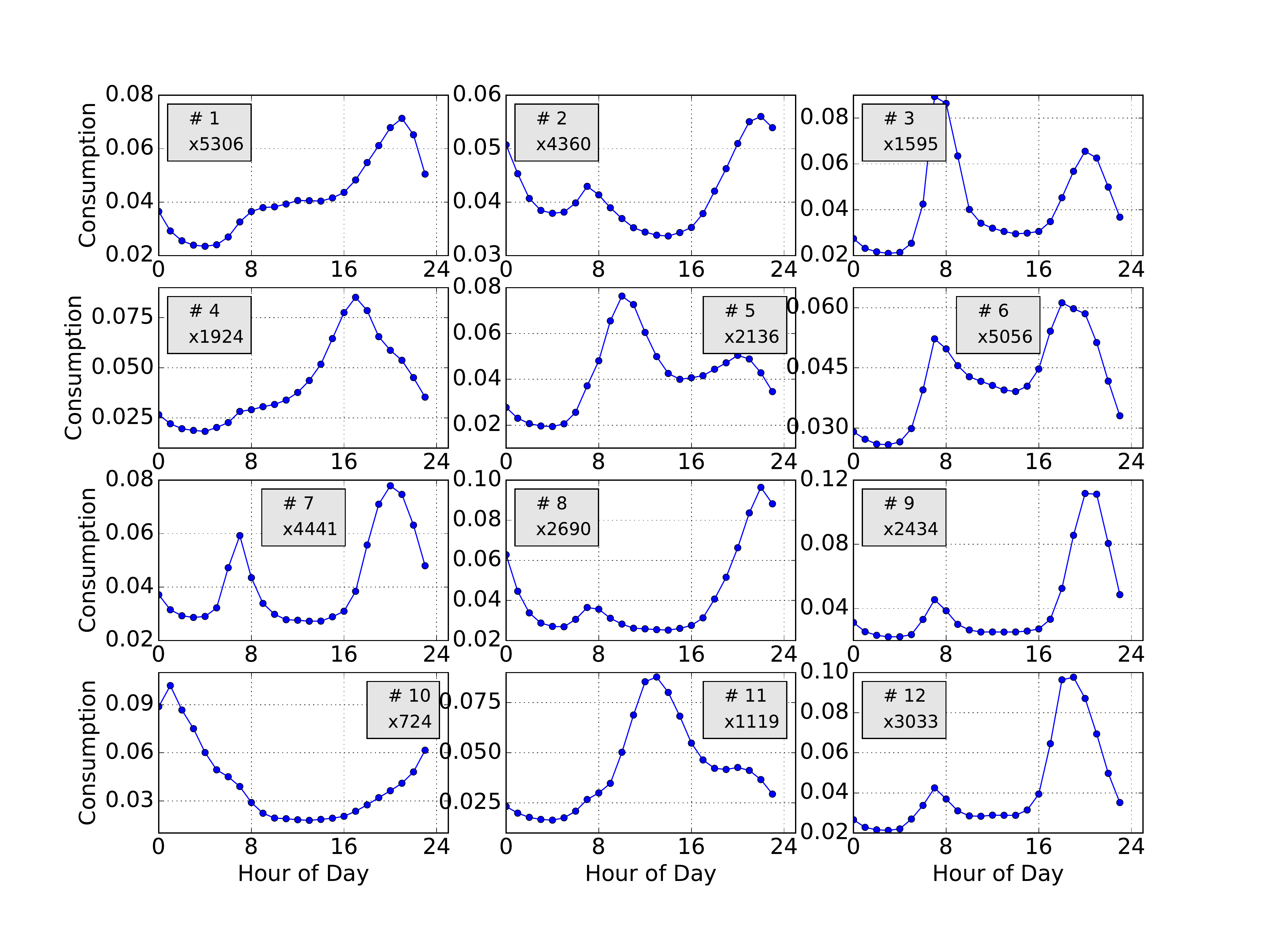}
\vspace*{-0.65cm}
\caption{Characteristic Load Shapes Identified with $k$-Means, $k=12$}
\label{fig:twelve_load_shapes}
\end{figure}

Figure \ref{fig:twelve_load_shapes} shows the 12 characteristic centroids and the number of load shapes that belong to the respective centroid.
Similar to \cite{Kwac:2014aa} and \cite{Smith:2012aa}, we can characterize different habits of users, such as users with a
\begin{itemize}
\item Morning + evening peak (\#2, \#3, \#6, \#7, \#9, \#12)
\item Daytime peak (\#5, \#11)
\item Night peak (\#8, \#10)
\item Evening peak (\#1, \#4)
\end{itemize}

\subsection{Entropies and P-Values}
\label{sec:Entropies_p_values}
\begin{figure*}[hbtp]
\subfloat[]{\includegraphics[width=1.0\textwidth]{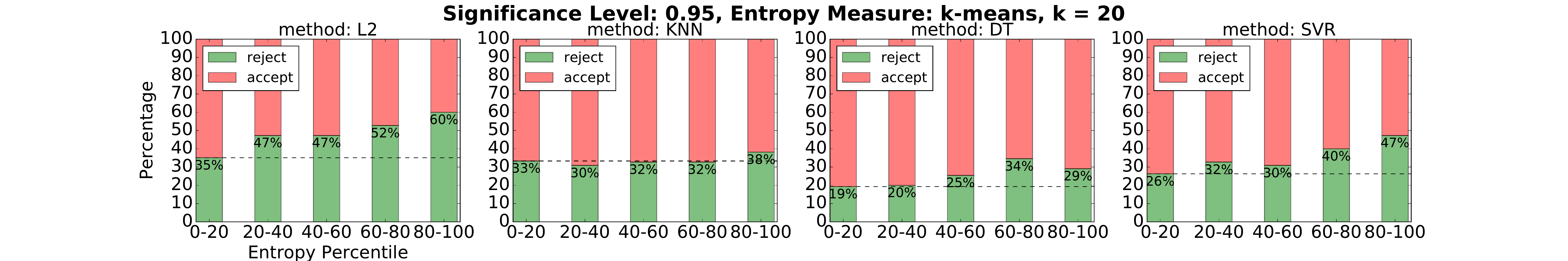}} \hfill \vspace*{-0.65cm}
\subfloat
[]{\includegraphics[width=1.0\textwidth]{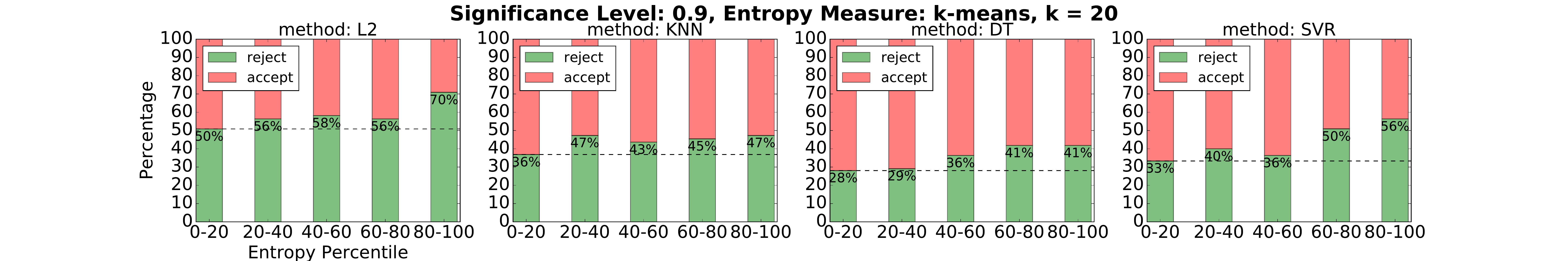}} \hfill \vspace*{-0.65cm}
\subfloat
[]{\includegraphics[width=1.0\textwidth]{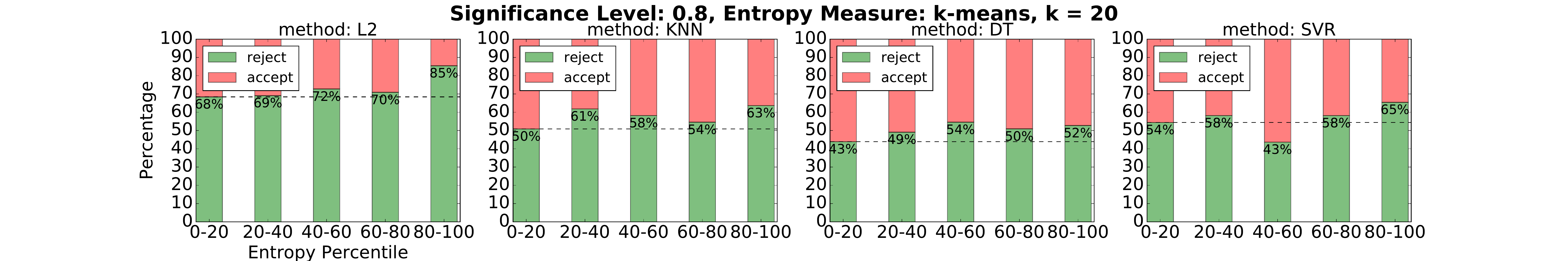}} \hfill \vspace*{-0.01cm}
\caption{Percentage of Rejected / Accepted Nulls for $k$-Means, 20 Clusters and Different Significance Levels $(1-p)$: Top: 0.95, Middle: 0.90, Bottom: 0.80}
\label{fig:pctg_acc_rej_20kmeans}
\vspace*{-0.2cm}
\end{figure*}

Figure \ref{fig:pctg_acc_rej_20kmeans} depicts bar charts for the percentage of accepted and rejected Null Hypotheses as defined in \eqref{eq:null_hypothesis_samples} for different significance levels and forecasting methods, sorted by entropy percentiles computed with \eqref{eq:classical_entropy_def} for $k=20$. For all significance levels, the percentage of rejections tends to increase as entropy increases. Under the assumption of a correctly specified model and in the absence of confounding variables, this suggests that users with higher variability in their consumption tend to have a lower consumption during DR events than those with lower variability in their consumption. Figure \ref{fig:pctg_acc_rej_20kmeans} shows a similar trend for $k$-means with 6 or 12 centroids as well as the standard deviation \eqref{eq:hourly_standard_dev} as entropy criteria.
An interesting observation is the tendency towards higher rejection rates for the linear regression models (OLS, L1, L2) compared to the nonparametric ones. This can be explained by the downward biased estimates of SVR and KNN, which reduce the estimated location parameter shifts $\hat{\Delta}$. A lower estimated location shift results in a smaller test statistic $U$, which then correlates with fewer rejected nulls in expectation.

\section{Conclusion}
\label{sec:Conclusion}
We analyzed Machine Learning methods for predicting residential energy consumption, and used these in conjunction with a non-parametric hypothesis test to estimate the flexibility in users' consumption during peak hours. We presented two entropy criteria for the variability of individual household consumption and identified a positive correlation between their inherent variability and the magnitude of the non-experimental estimates of reductions during DR periods.

The covariates used in our approach proved to yield satisfactory prediction results, and an improved choice of training features will only improve the forecasting accuracy, but not change our findings qualitatively. Further improvements can be achieved by incorporating a larger data set with more households and using more refined clustering methods. 

The effect of biased forecasts on the estimated DR reductions highlights the need for a more careful evaluation of the employed prediction methods, an issue that we are currently exploring. Due to the non-experimental nature of our estimates, in order to make claims about being able to identify the causal effects of DR interventions, our methods will need to be benchmarked against a randomized experiment.



\bibliographystyle{IEEEtran}
\bibliography{bibliography}


\end{document}